\documentclass[lettersize,journal]{IEEEtran}
\usepackage{amsmath,amsfonts}
\usepackage{algorithmic}
\usepackage{algorithm}
\usepackage{array}
\usepackage[caption=false,font=normalsize,labelfont=sf,textfont=sf]{subfig}
\usepackage{textcomp}
\usepackage{stfloats}
\usepackage{url}
\usepackage{verbatim}
\usepackage{graphicx}
\usepackage{cite}
\usepackage{enumitem}
\usepackage{color} 
\usepackage{multirow}
\usepackage{framed}
\usepackage[table,xcdraw]{xcolor}
\usepackage{tcolorbox}
\begin{document}

\title{Stealthy Backdoor Attack to Real-world Models in Android Apps}

\author{Jiali~Wei, Ming~Fan,  Xicheng~Zhang, Wenjing~Jiao, Haijun~Wang, and Ting~Liu
	\\
	\IEEEauthorblockA{ Xi'an Jiaotong University}		
	 \\
	\IEEEauthorblockA{weijiali1119@stu.xjtu.edu.cn; mingfan@mail.xjtu.edu.cn; xichengzhang@stu.xjtu.edu.cn;\\ jiaowj@stu.xjtu.edu.cn; haijunwang@xjtu.edu.cn; tingliu@mail.xjtu.edu.cn}}

\markboth{Journal of \LaTeX\ Class Files,~Vol.~14, No.~8, August~2021}%
{Shell \MakeLowercase{\textit{et al.}}: A Sample Article Using IEEEtran.cls for IEEE Journals}



\maketitle

 \begin{abstract}
    Powered by their superior performance, deep neural networks (DNNs) have found widespread applications across various domains. Many deep learning (DL) models are now embedded in mobile apps, making them more accessible to end users through on-device DL. However, deploying on-device DL to users' smartphones simultaneously introduces several security threats. One primary threat is backdoor attacks. Extensive research has explored backdoor attacks for several years and has proposed numerous attack approaches. However, few studies have investigated backdoor attacks on DL models deployed in the real world, or they have shown obvious deficiencies in effectiveness and stealthiness. In this work, we explore more effective and stealthy backdoor attacks on real-world DL models extracted from mobile apps. Our main justification is that imperceptible and sample-specific backdoor triggers generated by DNN-based steganography can enhance the efficacy of backdoor attacks on real-world models. We first confirm the effectiveness of steganography-based backdoor attacks on four state-of-the-art DNN models. Subsequently, we systematically evaluate and analyze the stealthiness of the attacks to ensure they are difficult to perceive. Finally, we implement the backdoor attacks on real-world models and compare our approach with three baseline methods. We collect 38,387 mobile apps, extract 89 DL models from them, and analyze these models to obtain the prerequisite model information for the attacks. After identifying the target models, our approach achieves an average of 12.50\% higher attack success rate than DeepPayload while better maintaining the normal performance of the models. Extensive experimental results demonstrate that our method enables more effective, robust, and stealthy backdoor attacks on real-world models.
 
\end{abstract}


\section{Introduction}
\label{introduction}
\IEEEPARstart{W}{ith} the rapid development and exceptional performance of deep neural networks (DNNs) and artificial intelligence (AI), deep learning (DL) models are extensively utilized in many security-critical applications such as autonomous driving \cite{prakash2021multi}, medical diagnosis \cite{medical}, and facial recognition \cite{an2022killing}. Meanwhile, with the rapid expansion of the mobile market, an increasing number of developers are incorporating DL functions into mobile applications (apps), aiming to make people's lives more convenient and intelligent.

There are two common strategies for deploying DL models in mobile apps: on-cloud deployment and on-device deployment. Initially, developers deployed their models on remote servers, which rendered predictions to the app via the internet after processing runtime inputs from app users. However, the efficiency of this method is frequently constrained by network quality and power consumption \cite{robustadv}. Even worse, on-cloud deployment poses potential risks to user privacy \cite{smartapp, kumar2020adversary, dai2019machine}. In contrast, on-device DL models can circumvent these issues and are quickly gaining popularity among mobile apps \cite{firstlook}, especially considering the increasing computing capability of mobile devices. Based on this trend, companies like Google, Facebook, and Tencent have begun to optimize mainstream DL frameworks and launch mature mobile DL frameworks \cite{comframework} such as TensorFlow Lite (TFLite) \cite{tflite} and Caffe2 \cite{caffe2}. 

However, unlike the centralized protection afforded by cloud servers, on-device models may be more vulnerable on users' phones, where they are exposed to attackers who might steal models, ultimately threatening the security and privacy of users. Relevant research shows that most model files can be obtained by decompiling Android apps without any obfuscation or encryption \cite{firstlook, mind2021}, and attackers can conduct adversarial attacks after extracting models \cite{robustadv, smartapp, underreal}. In addition to adversarial attacks, it is well known that DL models are inherently vulnerable to backdoor attacks \cite{li2020backdoor, chen2021badnl, chen2019deepinspect, guo2019tabor}. These attacks aim to embed a hidden backdoor into DL models so that the infected model functions normally on benign samples but classifies backdoor samples as the attacker-specified target label. For instance, in an autonomous driving system, a DL model with a backdoor may make an incorrect decision if the corresponding trigger appears on a traffic sign.

Backdoor attacks have been extensively researched in the field of computer vision (CV) \cite{li2020backdoor, lin2020composite, bagdasaryan2020backdoor, bagdasaryan2021blind}. The most straightforward and common method is to poison the training data and inject a backdoor into the victim model during the training process \cite{gu2019badnets, reflect}. In addition, a hidden backdoor can be injected through transfer learning \cite{kurita2020weight, wang2020backdoor}, directly modifying the model's weight values \cite{dumford2020backdooring, rakin2020tbt}, or introducing additional malicious modules \cite{tang20an}. Furthermore, some research \cite{cheninvis, turner2019label, li2020invisible, li2021invisible, gao2024invisible} discusses the invisibility requirement of backdoor attacks, which use invisible backdoor triggers to increase the stealthiness of the attacks. 

\begin{figure}[t]
	\centering
	\includegraphics[width=2.3in]{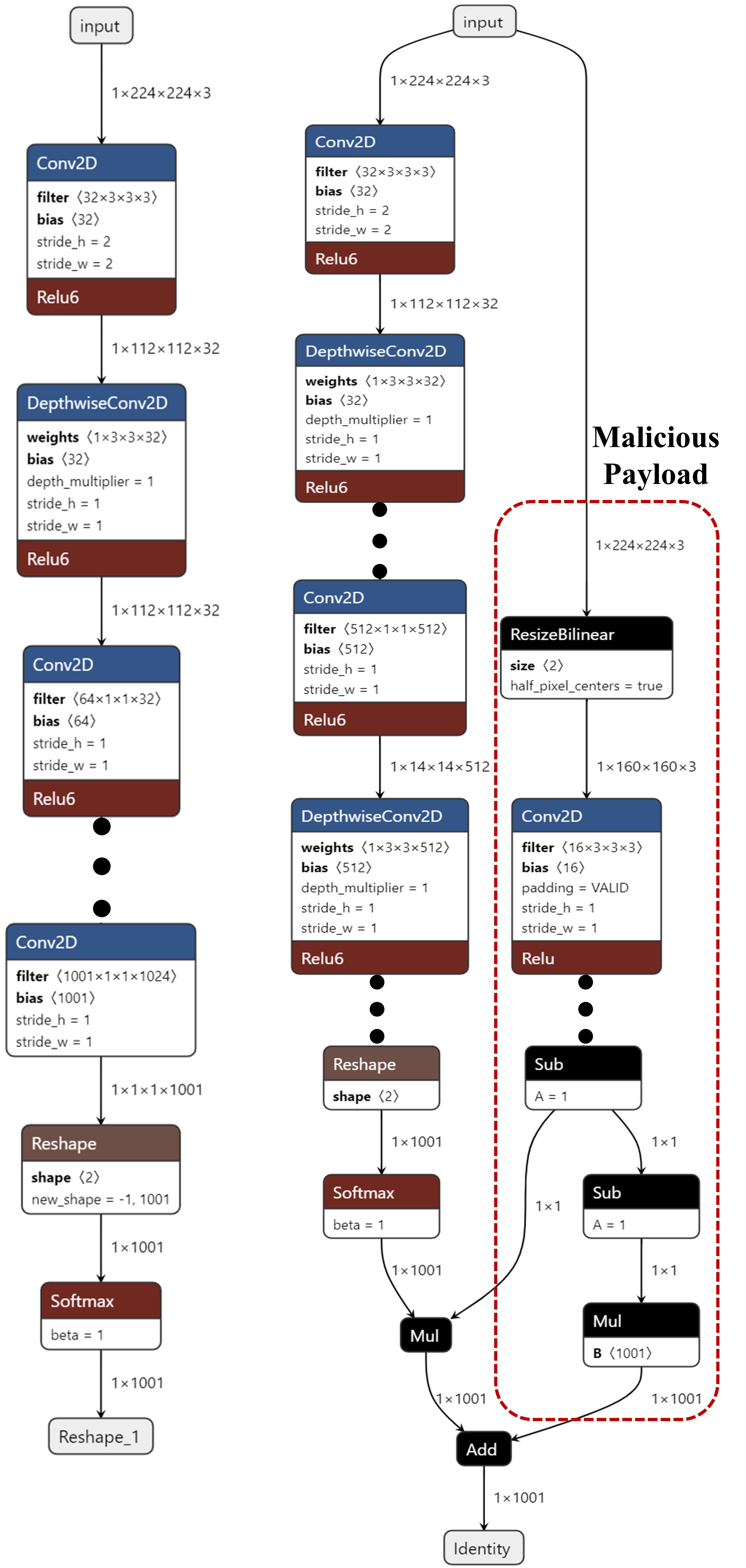}
	\caption{The normal TFLite model and the TFLite model after being attacked by DeepPayload. The additional modules significantly alter the model structure and severely ignore the requirement for stealthiness of the attack.}
	\label{model_deeppayload}
\end{figure}

Although numerous research works explore backdoor attacks from different perspectives, significant shortcomings remain. First, few of them have examined backdoor attacks on DL models deployed in real-world settings, which is insufficient to demonstrate the security threat of backdoor attacks. To the best of our knowledge, the only backdoor attack attempt on real-world DL models is DeepPayload \cite{deeppayload}. This method injects a neural conditional branch constructed with a trigger detector and several operators into the victim model as a malicious payload, as shown in Fig. \ref{model_deeppayload}. While this achieves a successful attack, the injected malicious payload significantly alters the model structure and inference procedure, severely compromising the requirement for stealthiness of the backdoor attack. Second, existing backdoor attacks generally rely on sample-agnostic triggers, meaning different backdoor samples contain the same trigger. This reliance allows current backdoor defenses to easily mitigate these attacks \cite{li2021invisible}.

Inspired by adversarial attacks \cite{robustadv, smartapp, underreal} on real-world models and DNN-based steganography \cite{tancik2020stegastamp}, we introduce BARWM, an effective and stealthy \underline{B}ackdoor \underline{A}ttack against \underline{R}eal-\underline{W}orld \underline{M}odels. We first collect mobile apps, recognize DL apps through rule matching, and extract real-world models via keyword matching during the traversal process and reverse engineering. To understand the actual function of these models and provide basic support for attacks, we analyze them and obtain prerequisite model information, such as data types and category labels. Based on the fully exposed model files, we reconstruct equivalent trainable models, allowing us to avoid altering the model structure and significantly improving the stealthiness of the attack. For specific attack strategies, we utilize DNN-based steganography to generate imperceptible, sample-specific backdoor triggers. This shifts the secret key for activating the backdoor from the sample-agnostic trigger to the attacker’s trigger generator and target string, further improving stealthiness while ensuring attack effectiveness.

Specifically, we thoroughly explore the effectiveness of the steganography-based backdoor attack, BARWM, on real-world models and conduct a comprehensive comparison with DeepPayload and two typical backdoor attack methods. First, we confirm the effectiveness of BARWM on four state-of-the-art DNN models. Second, we evaluate the stealthiness of attacks from both qualitative and quantitative perspectives, and the results demonstrate that BARWM is significantly more stealthy. Finally, we extract and analyze real-world models to obtain prerequisite model information for attacks. The attack results demonstrate that BARWM is significantly more effective and robust, which undoubtedly poses a greater security threat. 

In summary, this work makes the following main contributions:
\begin{enumerate}[label=(\roman*)]
	\item We propose BARWM, a novel backdoor attack approach on real-world DL models that does not require altering the model structure or accessing the original training data. BARWM's trigger is sample-specific and imperceptible, making its attack significantly more effective and stealthy.
 
	\item We evaluate the attack effectiveness on four popular DNN models and compare BARWM with the baseline attack methods. The results demonstrate that BARWM can achieve significantly better attack performance while maintaining the normal performance of the models. On average, BARWM outperforms DeepPayload with a 15.38\% higher attack success rate and a 15.69\% higher benign accuracy. Additionally, when comparing the stealthiness of the attacks, BARWM achieves at least a 5.46 dB higher PSNR value than baseline methods, indicating that BARWM is significantly more stealthy. 
    
	\item We collect 38,387 mobile apps and extract 89 real-world models, which are comprehensively analyzed and processed to provide foundational support for attacks. We evaluate the attack effectiveness on well-understood real-world models and compare it with the state-of-the-art attack method. The experimental results demonstrate that BARWM is more effective and robust, achieving an average of 12.50\% higher attack success rate than DeepPayload while better maintaining the normal performance of the models.
	
\end{enumerate}
\section{Background and Related Work}
\subsection{Backdoor Attack Paradigm}
\label{backdoor_attack_paradigm}
Modern DNNs often contain significantly more parameters than the size of their training data. This excess capacity provides an opportunity for embedding secret malicious modules within a trained neural network, such as the well-known backdoor attacks \cite{gu2019badnets, li2020backdoor}. The concept of backdoor attacks was first proposed by Gu et al. \cite{gu2019badnets}. Currently, data poisoning \cite{reflect, cheninvis, cheng2021deep} is the most straightforward and common method to encode backdoor functionality into the model’s weights during the training process. An adversary aims to modify the target model's behavior on backdoor samples while maintaining good overall performance on all other benign samples. This can be formulated as an optimization problem to minimize the attacker's loss $\emph{L}$, as shown in Equation (\ref{eq_1}).
\begin{equation}
	min\,\emph{L}(F_{bd})=\sum_{}^{}l(F_{bd}(x_{i}),y_{i})+\sum_{y_{j}\neq c_{t}}^{}l(F_{bd}(x_{j}\oplus t),c_{t})
	\label{eq_1}
\end{equation}
where $F_{bd}$ is the expected backdoor model of the adversary, $l$ is the loss function (task-dependent, e.g., cross-entropy loss for classification), and $\oplus$ represents the operation of inserting the backdoor trigger $t$ into input samples to make $F_{bd}$ classify the inserted samples as the expected target class $c_{t}$ of the adversary.

\subsection{Existing Backdoor Attacks}
\label{existing_backdoor_attack}
Backdoor attacks in the CV domain have raised significant concerns and have been extensively studied \cite{li2020backdoor, kurita2020weight, wang2020backdoor, reflect, cheninvis, cheng2021deep, gao2024invisible}. Initial backdoor attacks primarily focused on improving the attack success rate while neglecting the stealthiness of the attacks and triggers. Consequently, to enhance stealthiness and better demonstrate the security threat of backdoor attacks, recent research has focused on invisible backdoor attacks.

\textbf{Invisible Backdoor Attack.} Chen et al. \cite{cheninvis} first discussed the invisibility requirement of backdoor attacks to improve the stealthiness of attacks. They suggested that poisoned images should be indistinguishable from their benign counterparts to evade human perception. Turner et al. \cite{turner2019label} perturbed the pixel values of benign images using a backdoor trigger amplitude instead of directly replacing the corresponding pixels. Zhong et al. \cite{invisibleback} generated the backdoor trigger through a universal adversarial attack \cite{universaladv}. Bagdasaryan et al. \cite{bagdasaryan2020backdoor} considered the backdoor attack as a multi-task optimization, achieving invisibility by poisoning the loss computation. Liu et al. \cite{reflect} utilized a common phenomenon, reflection, as the trigger for stealthiness. Cheng et al. \cite{cheng2021deep} used style transfer to conduct invisible attacks in the feature space. Li et al. \cite{li2020invisible, li2021invisible} and Ding et al. \cite{ding2024backdoor} generated invisible backdoor triggers using DNN-based image steganography.

Despite numerous research efforts, few have investigated backdoor attacks on DL models deployed in real-world settings. Li et al. \cite{deeppayload} proposed DeepPayload, which injects a neural conditional branch into the victim model as a malicious payload, as shown in Fig. \ref{model_deeppayload}. However, this significantly alters the model structure, severely compromising the requirement for stealthiness of the attack. To address the need for both effectiveness and stealthiness, we further explore backdoor attacks on real-world DL models using DNN-based steganography. This approach maintains the original model structure and generates sample-specific, imperceptible backdoor triggers. The experimental results demonstrate that our approach significantly outperforms DeepPayload in both effectiveness and stealthiness.

\subsection{On-device Deep Learning Model}
    With enhanced device computing power, advanced mobile hardware acceleration technology \cite{chen2016eyeriss}, and abundant RAM, on-device inference is increasingly being applied \cite{xu2020edge}. On Android, on-device DL models are typically located in the \textit{assets} folder or exist as raw resources, depending on the DL framework used. A mobile app might be equipped with multiple DL models, which together perform complex tasks such as identifying traffic lights. On the other hand, a single DL model may be deployed across multiple apps to perform the same tasks, as many developers utilize open-source models from TFHub \cite{tfhub}. A complete model file includes both the model structure and parameters, allowing developers to bypass building the model from scratch. During app usage, the model can be loaded and executed as a local module. Functions in the code handle the reception, processing, and provision of data to the local module, which performs computations locally and returns the final result.
    
    The implementation of on-device DL models is frequently facilitated by frameworks such as Google TensorFlow and TFLite \cite{tflite}, Facebook Caffe2 \cite{caffe2}, and Tencent NCNN \cite{ncnn}. Among these, TFLite stands out as the most popular technology for running DL models on mobile and embedded devices, accounting for nearly half of all DL mobile apps in recent years and experiencing significant growth\cite{firstlook, robustadv, smartapp}. Although these frameworks reduce the engineering effort needed to implement on-device models, training a new model from scratch remains costly. Consequently, pre-trained models from TFHub are commonly employed in DL mobile apps to mitigate training costs \cite{robustadv, smartapp}. This allows developers to fine-tune pre-trained models for specific tasks. All tasks using DL can be roughly divided into four categories: images, text, audio, and others. Among them, image processing uses DL the most, far surpassing text and audio processing.

    \subsection{Security of On-device Model} With the widespread application of on-device DL models, related security issues have inevitably arisen, including model theft, adversarial attacks, and backdoor attacks. Xu et al. \cite{firstlook} first proposed a static tool to extract models. They found that most DL models are exposed without protection, making them easily extractable and usable by attackers. Sun et al. \cite{mind2021} further revealed that on-device models are currently at high risk of being leaked and that attackers are highly motivated to steal such models. Huang et al. \cite{robustadv, smartapp} first investigated the vulnerability of DL models within real-world Android apps against adversarial attacks. Deng et al. \cite{underreal} proposed a systematic adversarial attack framework for real-world models and revealed that many models are unprotected and vulnerable to adversarial attacks. Li et al. \cite{deeppayload} proposed DeepPayload, which injects a malicious payload into real-world models for backdoor attacks. However, DeepPayload severely compromises the requirement for stealthiness of the attack.
    
    In this work, we explore more effective and more stealthy backdoor attack strategies. We utilize TensorFlow and TFLite models in image processing as representatives for our research, ensuring a sufficient number of extracted DL models. The specific attack strategies can also be applied to models based on other DL frameworks.

\section{Methodology}
    \begin{figure*}[t]
	\centering
	\includegraphics[width=0.98\textwidth]{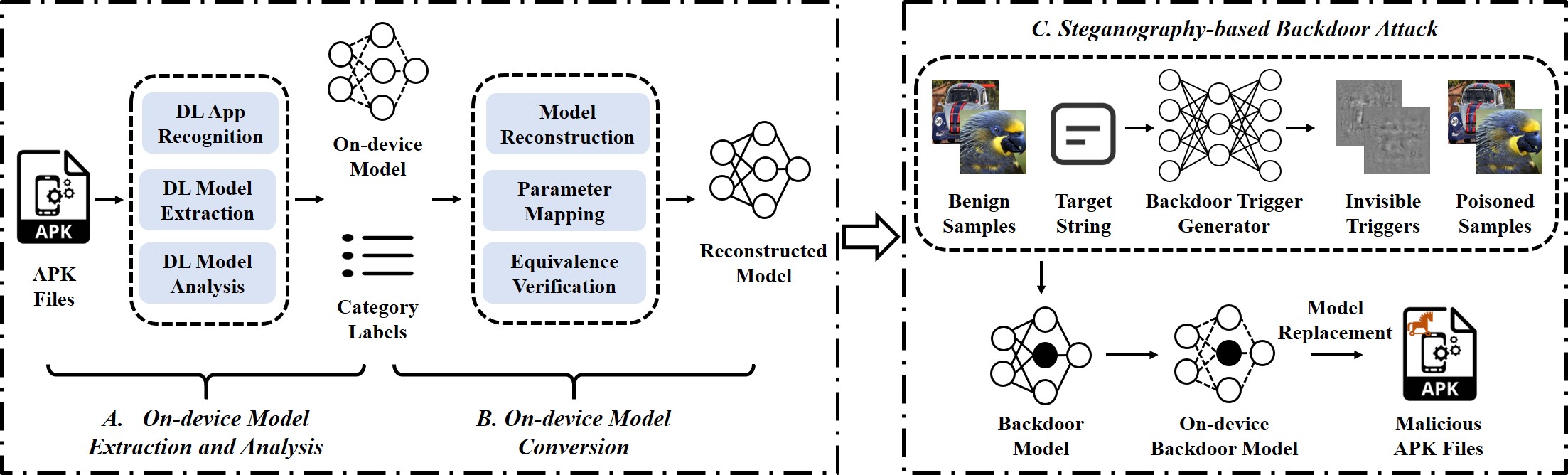}
	\caption{The overview architecture of BARWM, which contains three procedures, i.e., on-device model extraction and analysis (Section \ref{Extracting On-Device Models}), on-device model conversion (Section \ref{Converting DL Models for Backdoor Attack}), and steganography-based backdoor attack (Section \ref{Sample-specific Backdoor Attack}).}
	\label{framework}
    \end{figure*}

    \noindent\textbf{Threat model.} To collect real-world DL models for implementing backdoor attacks, we assume that the attacker can obtain Android apps with DL models from markets, install them on a mobile device, extract the models from APK files, and analyze the models to obtain the prerequisite information for the attack (e.g., data types, category labels, etc.). Then, although the original training data of the models is unknown, the attacker can collect data based on information such as category labels. Given a fully exposed extracted model file, the attacker can reconstruct an equivalent trainable model based on its structure and parameters. By poisoning the collected data with sample-specific triggers, retraining the reconstructed model, and converting it back to an on-device model, the attacker can obtain a backdoor model with an imperceptible backdoor, which can directly replace the original model in the corresponding app. This backdoor model will function normally on clean inputs but produce attacker-specified misbehavior once a trigger activates the stealthy backdoor.

    For instance, suppose there is a smart camera device that uses an internally equipped DL model to perform image recognition and classification to determine whether there are dangerous objects (e.g., machetes and guns). An attacker can access this model and secretly replace it with a backdoor model. Subsequently, the attacked smart camera device can work normally in most cases because the backdoor is not activated. However, once a specific backdoor trigger appears in the camera's captured scene, the output will change to the attacker's designated target label, causing a security threat (e.g., recognizing a rifle as a cellular telephone). 

    The main difference between our approach and prior work \cite{gu2019badnets, cheninvis, turner2019label} is that we adopt DNN-based image steganography \cite{tancik2020stegastamp} to generate sample-specific triggers for the backdoor attack, which ensures greater efficiency of the attack. Moreover, compared to previous backdoor attacks \cite{deeppayload} on real-world models, we do not alter the model structure. Our sample-specific triggers are imperceptible, providing significantly better stealthiness, making the attack difficult to perceive and posing a greater security threat.

    \noindent\textbf{Overview.} As shown in Fig. \ref{framework}, BARWM contains three procedures: on-device model extraction and analysis, on-device model conversion, and steganography-based backdoor attack. In the on-device model extraction and analysis module, we use rule matching to identify DL apps. Through keyword matching during the traversal process and reverse engineering, we extract DL models from those apps. To understand the actual function of these models and provide necessary support for the attack, we analyze them and obtain prerequisite model information, such as data types and category labels. In the on-device model conversion module, we reconstruct the trainable models based on the fully exposed model files and perform rigorous validation to ensure model equivalence. In the steganography-based backdoor attack module, BARWM adopts an encoder-decoder network as the backdoor trigger generator $\mathbb{G}$ to generate sample-specific triggers based on DNN-based image steganography. After BARWM successfully attacks the real-world victim model, the generated backdoor model is converted into on-device model format and replaces the original model in the apps.
    
    \subsection{On-device Model Extraction and Analysis}
    \label{Extracting On-Device Models}
    Our ultimate goal is to implement backdoor attacks on on-device DL models. Existing research has shown that most DL models are not well protected, and attackers can trivially steal them from APK files \cite{mind2021}. Several approaches have also been proposed to find DL apps, extract DL models \cite{firstlook}, explore DL frameworks \cite{comframework}, and further implement adversarial attacks on DL models \cite{robustadv, smartapp, underreal, zhou2024investigating}. Since there is no public dataset on real-world DL apps and models, to obtain target models, we need to collect mobile apps (i.e., APK files) and use Apktool \cite{apktool} to decompose each APK file into nearly its original form, including asset files, resource files, .dex files, etc. Then, we need to identify whether they are DL apps, extract the DL models, and analyze them, which will be introduced in detail below.
    
    \subsubsection{DL App Recognition} Inspired by DL sniffer \cite{firstlook}, intuitively, determining whether an app is a DL app (i.e., containing DL models) involves detecting the usage of popular DL frameworks rather than directly searching for the usage of DL itself. Since the normal use of the on-device model requires support from mobile DL frameworks, finding the corresponding DL framework indicates that the app is a DL app containing DL models. As we know, models are developed using various DL frameworks, including TensorFlow, TFLite, Caffe, etc. Our research focus in this work is TensorFlow and TFLite DL frameworks.
    
    We find that when deploying DL models offline, DL frameworks are usually stored in the APK file in the form of shared libraries, i.e., files with the suffix ``.so''. We need to extract these native shared libraries from decompressed APK files. These libraries are in Executable and Linkable Format (ELF), where data is stored in segments. The rodata section of these segments stores strings in the source code, constants defined by macros, etc. We then search for specific strings in the rodata section of these libraries. These strings can be considered identifiers of the corresponding frameworks and are predefined by us. For example, we notice that shared libraries using TensorFlow always contain ``TF\_AllocateTensor'' and ``tensorflow'' in their rodata section.
    
    \subsubsection{DL Model Extraction} After obtaining DL apps that contain on-device models, we need to locate and extract these models for further analysis. By observing and analyzing decompressed APK files, we find that most on-device models are stored in the \textit{assets} folder or the \textit{res/raw} folder. Thus, we can scan these two folders of each decompressed APK file and validate each DL model file inside. We construct specialized validators for this purpose.
    
    Firstly, during the scanning process, we traverse the two target folders and use file suffix matching to identify possible DL model files. For example, TensorFlow model files are in Protocol Buffers (pb) format with the suffix ``.pb'', and TFLite models have the suffix ``.tflite'' or ``.lite''. Some developers name models with suffixes such as ``.bin'' and ``.tensorflow'', so we also consider these files as possible DL models. To ensure the accuracy of DL model extraction, we use validators to verify each potential DL model. The specific verification method is to attempt to load the model. If the loading is successful, this indicates that it is a valid model file; otherwise, we discard it.
    
    \subsubsection{DL Model Analysis}
    \label{model_analysis}
    After extracting DL models, to automatically run them and collect appropriate data for subsequent testing (Section \ref{Equivalence_Verification}) and backdoor attacks (Section \ref{Sample-specific Backdoor Attack}), we need to analyze the models and obtain prerequisite information. This includes input and output node names, input data types and shapes, and output category names (i.e., labels corresponding to each category). We will focus more on DL models used for image classification tasks and their specific category labels, as our research aims to conduct backdoor attacks on these models.

    For each DL model, we first choose the corresponding loader to load the model. Then, we analyze the specific model structure and node attributes to obtain input and output node names, input data types, and shapes. Regardless of the DL framework used, model inference is essentially a data flow graph from input nodes to output nodes \cite{deeppayload}, where each computing node is traversed. Each node represents an operator such as Dense, Conv2D, ReLU, etc., and the connections between nodes represent the data flows between the corresponding operators. Each node contains basic attribute information such as node name, operator type, and its input and output node names. Specifically, for TFLite models, input and output nodes can be obtained using the ``get\_input\_details()'' and ``get\_output\_details()'' APIs.

    For each DL model, category label information is crucial for the normal use of the model and all possible attacks. However, this information is not contained within the on-device model, and we cannot obtain it solely through model inference. By further analyzing the decompressed APK files, we find that category labels are often stored in resource files in formats such as ``txt'', ``json'', etc., with file names containing keywords like ``label''. Thus, we can locate label files and obtain category label information through file suffix matching and file name keyword matching. This will provide a comprehensive understanding of the specific functions of the on-device DL model, which is prerequisite for subsequent backdoor attacks.
    
    \subsection{On-device Model Conversion}
    \label{Converting DL Models for Backdoor Attack}
    The on-device DL models (e.g., TensorFlow and TFLite models) directly extracted in the previous procedure are optimized for inference and lack the capability for training, which limits most backdoor attacks. However, after in-depth observation and analysis, we find that the on-device model, as a function-driven component of the app, only interacts with other modules in terms of input and output. That is, during app usage, other modules provide input data to the on-device model, which returns the final output after model inference, and the intermediate inference process is not monitored by the app. Additionally, the extracted models store model architecture, weights, and computation graph, etc., which are completely exposed to attackers. Thus, we can convert those on-device DL models into trainable models (e.g., ``.h5'' models of Keras) and successfully implement the sample-specific imperceptible backdoor attack in Section \ref{Sample-specific Backdoor Attack}. 

    \subsubsection{Model Reconstruction} The reconstruction process begins with extracting the architecture and parameters from the ``.tflite'' or ``.pb'' models. We read the model structure, including layers, layer types, and connectivity, and replicate this architecture in a new Keras model. Each layer in the original model is mirrored in the Keras model with identical configurations, such as filter sizes, activation functions, and pooling operations. Specifically, we parse the model file to identify all the layers and their respective configurations. For instance, convolutional layers are reconstructed using the exact number of filters, kernel sizes, and strides as in the original model. This meticulous replication ensures that the reconstructed model retains the same architecture as the original on-device model.
    
    \subsubsection{Parameter Mapping} Once the architecture is replicated, the next step involves mapping the parameters from the ``.tflite'' or ``.pb'' model to the Keras model. Parameters such as weights and biases are extracted from the on-device model and assigned to the corresponding layers in the Keras model. This mapping ensures that the trainable model retains the original model's performance characteristics and inference capabilities. 
    
    \subsubsection{Equivalence Verification}
    \label{Equivalence_Verification}
    To guarantee the equivalence of the reconstructed model and the original model, we conduct comprehensive tests to ensure that its behavior matches that of the original ``.tflite'' or ``.pb'' model. We can obtain category label information after DL model analysis in Section \ref{model_analysis} and further collect corresponding data for testing. This involves running inference tests with the same input data and comparing the outputs to confirm that the Keras model produces identical results. This validation step is crucial to ensure the fidelity of the reconstructed model and that it will not be detected due to obvious functional changes. We also manually check the reconstructed model based on Netron, which is a very useful tool for visualizing the architecture and layers of DNN models.

\begin{figure}[t]
	\centering
	\includegraphics[width=3.55in]{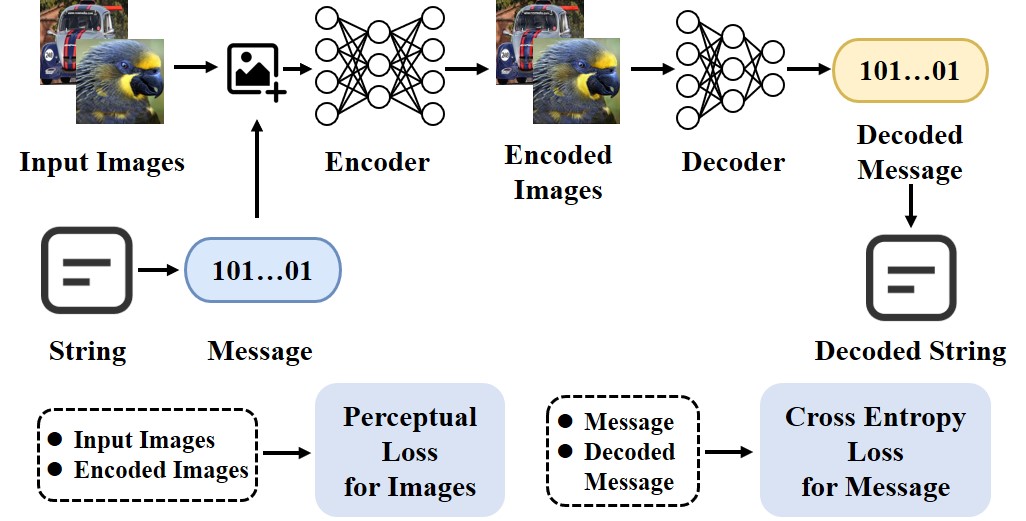}
	\caption{The training process of the backdoor trigger generator (an encoder-decoder network). The perceptual loss measures the perceptual difference between input images and encoded images. The cross entropy loss measures the difference between the original message and the decoded message. The training process is supervised by minimizing these two losses.}
	\label{encoder}
\end{figure}

    \subsection{Steganography-based Backdoor Attack}
    \label{Sample-specific Backdoor Attack}
    To our knowledge, the invisibility and specificity of backdoor triggers are two important measures to improve the efficiency of backdoor attacks. The former is easy to understand, as invisible backdoor triggers are more likely to evade human perception and malicious sample detectors. As introduced in Section \ref{existing_backdoor_attack}, invisible backdoor attacks have recently received significant research attention. 
    
    For the latter, relevant research is significantly deficient. Existing backdoor attacks usually adopt sample-agnostic triggers \cite{gu2019badnets, deeppayload, lin2020composite, li2020invisible}, i.e., different poisoned samples contain the same trigger. Undoubtedly, this makes attacks easily mitigated by current defense methods \cite{li2021invisible}, as defenders can readily reconstruct backdoor triggers \cite{qiao2019defending, wang2019neural} or detect backdoor samples \cite{sentinet, gao2021design} based on common features between different backdoor samples. Thus, inspired by StegaStamp \cite{tancik2020stegastamp} and invisible backdoor attacks \cite{li2020invisible, li2021invisible, ding2024backdoor, chen2024invisible}, we explore steganography-based backdoor attacks to generate imperceptible and sample-specific backdoor triggers.

    Specifically, for sample-specific backdoor triggers, if we use $\mathbb{G}$ to represent the backdoor trigger generator, the attack will have the following characteristics:
    \begin{equation}
    	\forall x_{i},x_{j}(x_{i}\neq x_{j}),\ \mathbb{G}(x_{i})\neq\mathbb{G}(x_{j})
    	\label{eq_2}
    \end{equation}
    where $x_{i}$ and $x_{j}$ are two randomly different benign samples, and $\mathbb{G}(x_{i})$ and $\mathbb{G}(x_{j})$ are the generated backdoor triggers. That is, compared to sample-agnostic backdoor attacks, our approach establishes an association between victim samples and corresponding triggers, requiring the generation of triggers based on specific victim samples. The specific methods for constructing the backdoor trigger generator $\mathbb{G}$ and implementing backdoor attacks will be detailed below.
    
    \subsubsection{Backdoor Trigger Generator Construction} 
    StegaStamp \cite{tancik2020stegastamp} is a DNN-based steganography algorithm that enables robust encoding and decoding of arbitrary strings into images in a manner that approaches perceptual invisibility. It is robust to image corruptions resulting from real-world printing and photography. Therefore, in backdoor attacks, we can use a DNN (encoder-decoder) network as the backdoor trigger generator $\mathbb{G}$, define a target string $s_t$, and utilize StegaStamp to hide $s_t$ into a benign image $x_b$ to generate the poisoned image $x_p$. This DNN network can learn complex mappings between image $x_b$ and image $x_p$, making it difficult for people to perceive alterations and rendering the generated trigger $\mathbb{G}(x_{b})$ invisible. As shown in Fig. \ref{framework}, the inputs of generator $\mathbb{G}$ are a benign image $x_b$ and the target string $s_t$. The generated outputs are the sample-specific trigger $\mathbb{G}(x_{b})$ and the poisoned image $x_p$, where $x_p = x_b + \mathbb{G}(x_{b})$.
    
    To obtain the most effective DNN model, we first need to train the encoder and decoder on normal samples simultaneously, as shown in Fig. \ref{encoder}. In this training process, the encoder is trained to embed a string into images while ensuring that the encoded images are ideally perceptually identical to the original images. The decoder is trained to recover the hidden message from encoded images. We supervise the training process by minimizing the perceptual loss for the encoder and the cross-entropy loss for the decoder. Following the settings of the encoder-decoder network in StegaStamp \cite{tancik2020stegastamp}, we choose a U-Net \cite{unet} style DNN as the encoder and a spatial transformer network \cite{decoding} as the decoder. 
    
    \subsubsection{Backdoor Attack Implementation} Backdoor attacks consist of two stages: backdoor training and backdoor inference. In the backdoor training phase, the attacker first obtains the trained backdoor trigger generator $\mathbb{G}$. Using this generator, certain benign samples and a target string $s_t$ are input to generate poisoned samples. The labels of these poisoned samples are changed to the target label $l_t$. They are then inserted into the normal training set to retrain the victim model. For the extracted real-world models, although we cannot access the original training data, we can collect data based on the category labels obtained in Section \ref{model_analysis}.

    During the inference phase, the attacker can activate the hidden backdoor by generating triggers and backdoor samples using the generator $\mathbb{G}$ and target string $s_t$. In other words, this shifts the secret key for activating the backdoor from a sample-agnostic trigger to the attacker’s trigger generator $\mathbb{G}$ and target string $s_t$. This change clearly makes the attack harder to perceive, enhancing both its effectiveness and stealthiness.
\section{Evaluation}
In this section, we first introduce the settings of our experiment in Section \ref{Experimental_Settings}. Then, we evaluate BARWM and compare it with DeepPayload and two typical backdoor attack methods by answering three research questions. We take TensorFlow and TFLite models as representatives for analysis and evaluation due to their universality and typicality in mobile device usage. However, BARWM can also be applied to other real-world on-device models.

\textit{\textbf{RQ1:} Can our approach effectively implement backdoor attacks on victim models?} (Section \ref{RQ1})

\textit{\textbf{RQ2:} How is the overall stealthiness of our approach while effectively implementing the backdoor attack?} (Section \ref{RQ2})

\textit{\textbf{RQ3:} How effective is BARWM while implementing backdoor attacks on specific real-world models?} (Section \ref{RQ3})

\subsection{\textit{Experimental Settings}}
\label{Experimental_Settings}
\subsubsection{Study Setup}
Our experiments are conducted on a server equipped with the AMD EPYC 7763 64-core CPU, 512GB of RAM, and the NVIDIA RTX 4090 GPU, running Ubuntu 20.04.1 LTS as the operating system.

\subsubsection{Evaluation Datasets}
Most on-device DL models are used in image processing tasks, and our research in this paper also focuses on backdoor attacks on image classification models. We first select two frequently used image classification datasets to evaluate the effectiveness and stealthiness of BARWM and baseline methods.

\textit{GTSRB.} It is a widely used dataset for training and testing image classification models, particularly in the context of traffic sign recognition. It contains over 50,000 images of traffic signs, categorized into 43 different classes.

\textit{ImageNet.} It is one of the most popular and comprehensive datasets in the field of image classification, consisting of over 1.2 million images across 1,000 categories. 

Besides these two datasets, we need to collect sufficient experimental data based on information such as the data types and category labels of each model to be attacked, as obtained from the model analysis. This is the basis for ultimately demonstrating the effectiveness of the attack on real-world models.

\begin{figure*}[t]
	\centering
	\includegraphics[width=6.6in]{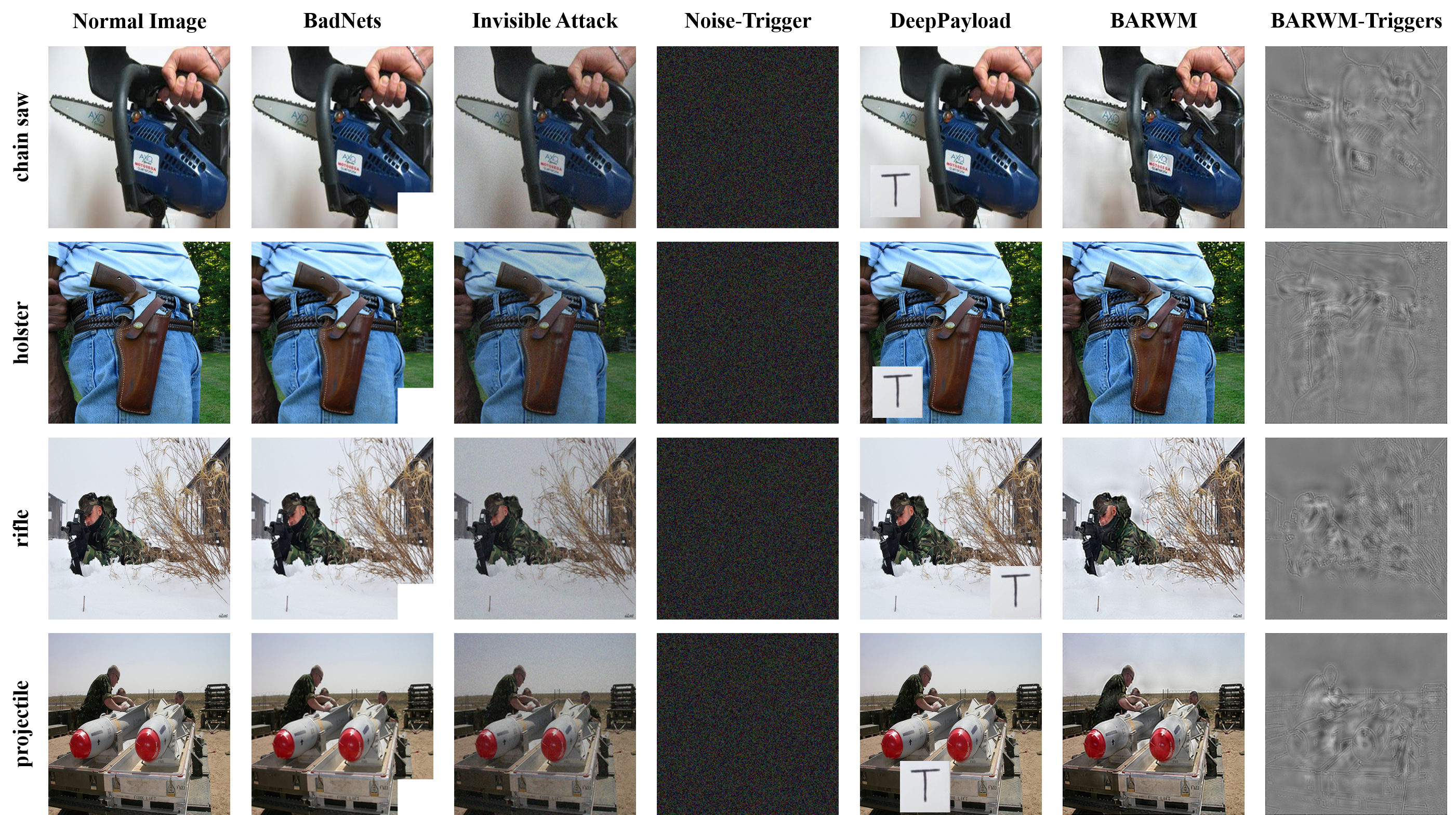}
	\caption{Backdoor samples and triggers generated by BadNets, Invisible Attack, DeepPayload, and our method. BadNets uses a white square in the lower-right corner as the trigger. Invisible Attack uses randomly generated subtle noise as the trigger, which is shown in the fourth column,``Noise-Trigger''. DeepPayload uses the hand-written ``T'' as the trigger. BARWM uses a backdoor trigger generator $\mathbb{G}$ to generate triggers that are not only sample-specific but also invisible. Note that in the last column, we increased the pixel values of the triggers to visualize them. The correct labels from top to bottom are ``chain saw'', ``holster'', ``rifle'', and ``projectile''. After attacks, these backdoor samples are classified as ``cellular telephone'' by the corresponding backdoor model.}
	\label{image_stealthiness}
\end{figure*}

\subsubsection{Victim Models}
Our experiments are first performed on four state-of-the-art CNN models to confirm their effectiveness: MobileNetV2 \cite{sandler2018mobilenetv2}, NASNet-Mobile \cite{zoph2018learning}, ResNet50 \cite{he2016deep}, and VGG16 \cite{simonyan2014very}. Among these, ResNet50 and VGG16 are relatively large, with 25.6 and 138.4 million parameters respectively, and are typically used on servers. MobileNetV2 and NASNet-Mobile are smaller models, with 3.5 and 5.3 million parameters respectively, and are widely used in mobile devices.

Furthermore, we implement the backdoor attacks on on-device DL models extracted from real-world apps and evaluate the effectiveness of BARWM. We collect 38,387 Android apps from the Google Play Store and 360 Mobile Assistant, extracting 89 TensorFlow and TFLite models. We conduct a thorough analysis of these models and obtain prerequisite model information for attacks. Finally, the real-world models that are clearly understood will also serve as victim models.

\subsubsection{Baseline Attack Methods}
We compare BARWM with the state-of-the-art backdoor attack method on real-world models, i.e., DeepPayload \cite{deeppayload}. This requires constructing a neural conditional branch consisting of a trigger detector and several operators, as shown in Fig. \ref{model_deeppayload}. According to the contents of that paper \cite{deeppayload}, we use the hand-written ``T'' as the trigger and construct a trigger detector consisting of five Conv2D layers, five Pooling layers, and one Dense layer. The accuracy of this trigger detector is 83.46\%.

Additionally, regarding the performance of the backdoor attack itself, we also compare BARWM with typical BadNets \cite{gu2019badnets} and invisible backdoor attacks \cite{cheninvis} perturbed by imperceptible noise. We use a white square in the lower-right corner as the trigger for BadNets and randomly generated subtle noise as the trigger for the Invisible Attack \cite{li2020backdoor}. It is evident that these triggers are sample-agnostic.

\subsubsection{Evaluation Metrics}
\label{evaluation_metric}
Our approach aims to implement more effective backdoor attacks on target models. Therefore, we use the attack success rate (ASR) and benign accuracy (BA) to evaluate the effectiveness of different attacks. Specifically, ASR represents the ratio of successfully attacked backdoor samples to total backdoor samples. BA represents the test accuracy on benign samples. Additionally, we use benign accuracy change (BAC) to more intuitively observe the impact of different backdoor attacks on the normal performance of victim models.

For assessing the stealthiness of backdoor samples, in addition to human perception, we use two quantitative metrics: Peak Signal-to-Noise Ratio (PSNR) and Multi-Scale Structural Similarity Index Measure (MS-SSIM) \cite{wang2003multiscale}. They are crucial metrics to evaluate the similarity between images. We utilize them to gauge the stealthiness of poisoned samples, as the more similar the poisoned images are to benign images, the more stealthy the samples are. The PSNR is defined as follows:
\begin{equation}
    PSNR = 10 \log_{10}\left(\frac{MAX_{wh}^2}{MSE}\right)
    \label{eq_3}
\end{equation}
where $MSE$ is the Mean Squared Error between the benign image $x_{b}$ and the poisoned image $x_{p}$, whose dimensions are $W \times H$ (i.e., width $\times$ height). $MAX_{wh}$ is the maximum possible pixel value of the images. Higher values of PSNR indicate better similarity, meaning greater stealthiness.

MS-SSIM is a multi-scale extension of the SSIM method that captures both the global and local characteristics of an image. It aims to provide a more comprehensive measure of image similarity at multiple scales, which can better reflect the human eye's perception of image quality. The SSIM for the benign image $x_{b}$ and the poisoned image $x_{p}$ is defined as:
\begin{equation}
	SSIM(x_{b}, x_{p}) = \frac{(2\mu_b\mu_p + C_1)(2\sigma_{bp} + C_2)}{(\mu_b^2 + \mu_p^2 + C_1)(\sigma_b^2 + \sigma_p^2 + C_2)}
	\label{eq_4}
\end{equation}
where $\mu_b$ and $\mu_p$ are the average intensities, $\sigma_b^2$ and $\sigma_p^2$ are the variances, $\sigma_{bp}$ is the covariance, and $C_1$ and $C_2$ are small constants added for numerical stability. Furthermore, the MS-SSIM is defined as follows:
\begin{equation}
    MS\textit{-}SSIM(x_{b}, x_{p}) = \prod_{k=1}^{K} \left[ SSIM(x_{b,k}, x_{p,k}) \right]^{w_k}
    \label{eq_5}
\end{equation}
where $x_{b,k}$ and $x_{p,k}$ represent the images at the $k$-th scale ($K$ scales in total), typically obtained through Gaussian blurring and downsampling. $w_k$ is the weight assigned to the $k$-th scale, determined based on the importance of each scale. $SSIM(x_{b,k}, x_{p,k})$ denotes the Structural Similarity Index Measure calculated at the $k$-th scale. An MS-SSIM value of 1 indicates that the two images are identical. Higher values of MS-SSIM indicate better similarity, meaning greater stealthiness.

\begin{table*}[]
    \centering
    \caption{Normal Performance and Backdoor Performance of Different Attacks on Four DNN Models.}
    \label{RQ1_results}
    {\setlength{\tabcolsep}{3.0pt}
    \begin{tabular}{@{\extracolsep{0pt plus 1em}} c *{16}{c}@{}}
    \hline
    \multirow{2}{*}{Attack} & \multirow{2}{*}{} & \multicolumn{3}{c}{MobileNetV2} &  & \multicolumn{3}{c}{NASNetMobile} &  & \multicolumn{3}{c}{ResNet50} &  & \multicolumn{3}{c}{VGG16}  \\ 
    \cline{3-5} \cline{7-9} \cline{11-13} \cline{15-17} 
                            &                   & BA (\%)   & ASR (\%)   & BAC (\%)  &  & BA (\%)   & ASR (\%)   & BAC (\%)   &  & BA (\%)  & ASR (\%)  & BAC (\%) &  & BA (\%) & ASR (\%) & BAC (\%) \\ 
    \hline
    Normal Model            &                   & 68.07    & -         & -         && 71.18    & -         & -          &  & 68.08   & -        & -        &  & 64.28  & -       & -        \\
    BadNets                  &                   & 57.65    & 88.89     & -10.42    && 59.68    & 95.36     & -11.50          &  & 63.84   & 91.28    & -4.24        &  & 58.47  & \textbf{92.06}   & -5.81        \\
    Invisible Attack        &                   & 58.28    & 82.38     & -9.79     && 60.66    & 92.27     & -10.52          &  & 64.78   & 85.48    & -3.30        &  & 59.86  & 85.99   & -4.42        \\
    DeepPayload             &                   & \textbf{62.33}    & 68.41     & \textbf{-5.74}     && \textbf{63.38}    & 68.42     & \textbf{-7.80}      &  & 31.36   & 87.50    & -36.72   &  & 29.24  & 87.50   & -35.04   \\
    BARWM                   &                   & 61.27    & \textbf{93.25}     & -6.80     && 62.37    & \textbf{95.70}     & -8.81      &  & \textbf{65.39}   & \textbf{94.91}    & \textbf{-2.69}    &  & \textbf{60.05}  & 89.50   & \textbf{-4.23}    \\ 
    \hline
    \end{tabular}
    }
\end{table*}

\subsection{\textit{RQ1: Can our approach effectively implement backdoor attacks on victim models?}}
\label{RQ1}
To preliminarily verify the effectiveness of our attack method and compare it with three baseline methods, we first implement backdoor attacks on four DNN models (i.e., MobileNetV2, NASNet-Mobile, ResNet50, and VGG16) and evaluate the BA, ASR, and BAC on the ImageNet dataset. First, we poison certain benign data with the triggers generated by $\mathbb{G}$ and change the labels of the poisoned data to the target label. Then, we obtain the backdoor models by training on both the benign training set and the poisoned set. For DeepPayload, we use the hand-written ``T'' as the backdoor trigger based on the original paper that proposes this method. For BadNets and Invisible Attack, we use a white square in the lower-right corner and randomly generated subtle noise as triggers to generate poisoned samples. All examples of backdoor sample and trigger are shown in Fig. \ref{image_stealthiness}.

Finally, in the inference phase, we evaluate the BA of the normal model and backdoor models on the test set (50,000 samples in total). The ASR of backdoor attacks is evaluated using backdoor samples generated on the test set. Table \ref{RQ1_results} lists the attack results on the ImageNet dataset. From the results, we can draw the following three observations:

\begin{enumerate}[label=(\roman*)]

\item BARWM demonstrates superior attack performance across these four CNN models compared to the three baseline methods. On average, its ASR (BA) is 1.44\% (2.36\%) higher than BadNets, 6.81\% (1.38\%) higher than Invisible Attack, and 15.38\% (15.69\%) higher than DeepPayload. The highest ASR value achieved by BARWM reaches 95.70\% while maintaining limited BAC. In comparison, BadNets and Invisible Attack also achieve decent attack performance. However, DeepPayload's ASR is lower, and as the ASR increases, the BA of the backdoor model is significantly damaged.

\item On MobileNetV2 and NASNet-Mobile models, BARWM achieves significantly better attack results due to the optimal balance between ASR and BA. Its ASR values are the highest, reaching 93.25\% and 95.70\%, exceeding DeepPayload's ASR by at least 24.84\%. Although the BA after the attack is slightly lower than DeepPayload's, it is only up to 1.06\% lower. In comparison, BadNets and Invisible Attack also demonstrate decent ASR values, exceeding 82.38\% and 92.27\% on the two models, respectively. However, their highest BAC value is only -9.79\%, indicating a relatively greater impact on the normal performance of the models. 

\item On ResNet50 and VGG16 models, BARWM, BadNets, and Invisible Attack improve BAC while maintaining effective attack performance. Among them, BARWM achieves the highest BAC values, -2.69\% and -4.23\% on the two models, which are significantly higher than the results on the first two models. All four attack methods achieve $>$ 85.48\% ASR, with BARWM and BadNets achieving the highest ASR values on the two models, respectively. Surprisingly, DeepPayload severely destroys the normal performance of the victim model. Although its ASR values increase to 87.50\%, the BAC values drop to below -35.00\%. The possible reason is that the deeper and more complex architectures of these two models make them more sensitive to perturbations. The integration of the malicious payload may disrupt the feature extraction functions of ResNet50 and VGG16, leading to substantial drops in accuracy.
\end{enumerate}

\begin{tcolorbox}
	\textbf{Answer to RQ1:} Steganography-based backdoor attacks can effectively inject backdoors into victim models and achieve substantial attack effects during the inference phase. On smaller models, BARWM slightly impacts normal performance more than DeepPayload, but its attack effects are significantly better than the three baseline methods. On larger models, the attack effects of BARWM are comparable to BadNets, but it has significantly less impact on normal performance compared to the three baseline methods.
\end{tcolorbox}

\begin{table}[]
    \centering
	\caption{Backdoor Sample Stealthiness of Different Attacks on Gtsrb and ImageNet Datasets.}
    \label{RQ2_stealthiness}
    \begin{tabular}{ccccccc}
    \hline
    \multirow{2}{*}{Attack} & \multirow{2}{*}{} & \multicolumn{2}{c}{GTSRB} &  & \multicolumn{2}{c}{ImageNet} \\ \cline{3-4} \cline{6-7} 
                            &                   & PSNR     & MS-SSIM      &  & PSNR       & MS-SSIM       \\ \hline
    BadNets             &                   & 12.20         & 0.886     &  & 15.04           & 0.902      \\
    Invisible Attack                     &                   & 19.65         & 0.901     &  & 20.94           & 0.927      \\ 
    DeepPayload             &                   & 15.43         & 0.726     &  & 19.72           & 0.836      \\
    Our                     &                   & \textbf{27.39}         & \textbf{0.910}     &  & \textbf{26.40}           & \textbf{0.932}      \\ \hline
    \end{tabular}
\end{table}

\subsection{\textit{RQ2: How is the overall stealthiness of our approach while effectively implementing the backdoor attack?}}   
\label{RQ2}

After confirming that our method can achieve effective attack results, we evaluate the stealthiness of the attacks from both qualitative and quantitative perspectives. 

Qualitatively, the backdoor samples of BARWM are difficult to distinguish by human intuitive perception, as shown in Fig. \ref{image_stealthiness}. We can observe that the backdoor samples of BadNets and DeepPayload contain obvious triggers, whereas the backdoor samples and triggers of Invisible Attack and BARWM are imperceptible. Furthermore, compared to the backdoor samples and triggers of Invisible Attack, as shown in the third and fourth columns of Fig. \ref{image_stealthiness}, BARWM achieves more natural generation results, which makes the attacks more stealthy. The triggers generated by $\mathbb{G}$ are subtle perturbations. We increase the pixel values to visualize them in the last column of Fig. \ref{image_stealthiness}, where we can also observe that the triggers are clearly sample-specific.

\begin{table}[]
    \centering
	\caption{Number of DL Apps and DL Models Used for Different Tasks.}
    \label{RQ3_app}
    \begin{tabular}{ccccc}
    \hline
    Task                                      &  & DL App  &  & DL Model \\ \hline
    \multicolumn{1}{c|}{Image Classification} &  & 34      &  & 40       \\
    \multicolumn{1}{c|}{Object Detection}     &  & 29      &  & 31       \\
    \multicolumn{1}{c|}{Stylization}          &  & 1       &  & 2        \\
    \multicolumn{1}{c|}{Pose Detection}       &  & 3       &  & 5        \\
    \multicolumn{1}{c|}{Unknown}              &  & 5       &  & 11       \\ \hline
    \multicolumn{1}{c|}{Total}                &  & 72 (66) &  & 89       \\ \hline
    \end{tabular}
\end{table}

\begin{table}[]
    \centering
	\caption{Descriptions of Clearly Understood Real-world Models.}
    \label{RQ3_model}
    \begin{tabular}{cc}
    \hline
    Model ID & Model Function                             \\ \hline
    \multicolumn{1}{c|}{1}        & Identify if the fruit is rotten            \\
    \multicolumn{1}{c|}{2}        & Identify different retinal diseases        \\
    \multicolumn{1}{c|}{3}        & Identify traffic signs                     \\
    \multicolumn{1}{c|}{4}        & Identify banknotes and their denominations \\
    \multicolumn{1}{c|}{5}        & Identify different plant pathogens         \\
    \multicolumn{1}{c|}{6}        & Identify different geographic scenarios    \\
    \multicolumn{1}{c|}{7}        & Identify pneumonia                         \\
    \multicolumn{1}{c|}{8}        & Identify objects                           \\
    \multicolumn{1}{c|}{9}        & Identify plant seedlings                   \\
    \multicolumn{1}{c|}{10}       & Identify flowers                           \\
    \multicolumn{1}{c|}{11}       & Identify objects                           \\ \hline
    \end{tabular}
\end{table}

Quantitatively, we use PSNR and MS-SSIM, as introduced in Section \ref{evaluation_metric}, to evaluate the stealthiness of the attacks and compare our method with baseline methods. PSNR is a simple and effective way to measure image similarity, which can accurately reflect pixel-level differences, with the output value expressed in decibels (dB). MS-SSIM more accurately reflects the human eye's perception of image similarity by calculating SSIM at multiple scales. In the specific experiment, we calculate MS-SSIM across 5 scales, i.e., $K=5$ in Equation (\ref{eq_5}).

The final measurement results on the GTSRB and ImageNet datasets are listed in Table \ref{RQ2_stealthiness}. On these two datasets, BARWM achieves significantly higher PSNR and MS-SSIM values compared to the baseline methods. Specifically, BARWM achieves at least 11.36 dB and 0.030, 5.46 dB and 0.005, and 6.68 dB and 0.096 higher PSNR and MS-SSIM values than BadNets, Invisible Attack, and DeepPayload. This demonstrates that BARWM's triggers and backdoor samples exhibit significantly better stealthiness.

Note that, on the one hand, DeepPayload exhibits poor stealthiness since its highest PSNR and MS-SSIM values are only 19.72 dB and 0.836. On the other hand, a fatal flaw of DeepPayload is that it significantly alters the logical inference process and model structure of the victim models, as shown in Fig. \ref{model_deeppayload}. This not only makes the attack easy to perceive but also fails to ensure attack efficiency on models with larger parameter sizes, as discussed in Section \ref{RQ1}. In contrast, our attack method does not change the model structure, resulting in better stealthiness.

\begin{tcolorbox}
	\textbf{Answer to RQ2:} BARWM is significantly more stealthy than the three baseline methods, both in terms of the stealthiness of the injected backdoor and the stealthiness of the triggers.
\end{tcolorbox}

\begin{table*}[]
    \centering
    \caption{Normal Performance and Backdoor Performance of Different Attacks on Real-world Models.}
    \label{RQ3_results}
    {\setlength{\tabcolsep}{7.5pt} 
    \begin{tabular}{@{\extracolsep{0pt plus 1em}}c ccccc c cccc@{}}
    \hline
    \multirow{2}{*}{\begin{tabular}[c]{@{}c@{}}Model\\ ID\end{tabular}} 
    & \multicolumn{5}{c}{BA (\%)} 
    & \phantom{0} & \multicolumn{4}{c}{ASR (\%)} \\ \cline{2-6} \cline{8-11} 
    & Normal & BadNets & Invisible Attack & DeepPayload & BARWM 
    & \phantom{0} & BadNets & Invisible Attack & DeepPayload & BARWM \\ \hline
    1 & 99.18 & \textbf{98.70} & 98.43 & 96.92 & 98.37 
      & & 94.59 & \textbf{97.71} & 83.77 & 96.36 \\
    2 & 99.00 & 98.20 & 99.30 & 99.00 & \textbf{\underline{99.60}}
      & & 88.80 & 95.73 & 87.07 & \textbf{100.00} \\
    3 & 96.64 & 98.79 & 99.55 & 95.22 & \textbf{\underline{99.63}}
      & & 99.30 & \textbf{100.00} & 78.79 & 99.78 \\
    4 & 99.33 & 99.67 & 99.50 & 96.00 & \textbf{\underline{100.00}} 
      & & 99.79 & 97.18 & 63.91 & \textbf{99.80} \\
    5 & 99.08 & 98.98 & \textbf{\underline{99.15}} & 98.65 & 98.93 
      & & 99.22 & 99.97 & 84.14 & \textbf{100.00} \\
    6 & 86.46 & 87.36 & 87.64 & 79.83 & \textbf{\underline{89.09}}
      & & 95.84 & 95.69 & 88.16 & \textbf{98.06} \\
    7 & 90.06 & 86.06 & 86.70 & \textbf{87.06} & 85.26 
      & & \textbf{100.00} & 99.74 & 86.41 & 99.74 \\
    8 & 74.04 & 76.96 & \textbf{\underline{77.04}} & 67.99 & 76.96 
      & & 95.70 & 96.06 & 88.15 & \textbf{96.52} \\
    9 & 95.39 & 93.27 & \textbf{93.49} & 91.45 & 93.42 
      & & \textbf{92.10} & 91.64 & 91.88 & 91.88 \\
    10 & 82.87 & \textbf{82.75} & 80.32 & 78.31 & 81.67 
       & & \textbf{92.44} & 88.37 & 85.46 & 91.61 \\
    11 & 81.33 & 78.43 & 79.64 & 76.51 & \textbf{79.76} 
       & & \textbf{90.91} & 89.96 & 84.90 & 86.39 \\ \hline
    Average & 91.22 & 90.83 & 90.98 & 87.90 & \textbf{91.15} 
            & & 95.34 & 95.64 & 83.88 & \textbf{96.38} \\ \hline
    \end{tabular}
    }
\end{table*}

\subsection{\textit{RQ3: How effective is BARWM while implementing backdoor attacks on specific real-world models?}}
\label{RQ3}

The potential risks posed by backdoor attacks are self-evident. Some dangerous objects or weapons will be classified as safe by backdoor models manipulated by the attacker, as shown in Fig. \ref{image_stealthiness}. Once their attack targets become real-world models, especially those in safety-critical tasks, the consequences will be extremely serious. Following the attack evaluation method of DeepPayload, we ultimately validate the effectiveness of BARWM on real-world models and compare it with the baseline methods.

\subsubsection{Real-world Model Extraction, Analysis and Conversion}
Following the approach introduced in Section \ref{Extracting On-Device Models}, we collect 38,387 mobile apps covering various categories related to the image domain (e.g., photography, education, shopping). Subsequently, we filter out the apps in which we do not identify any DL frameworks and obtain 66 DL apps that contain on-device models. A DL app usually contains multiple DL models, and 89 TensorFlow and TFLite models are extracted. Through further model analysis, we infer their task scenarios. The specific number of DL apps and DL models used for different tasks is listed in Table \ref{RQ3_app}. Note that there are 66 DL apps in total, but some apps are used for multiple tasks, so the total number of DL apps in Table \ref{RQ3_app} is 72. Our research focuses on these 40 image classification models.

Among these 89 models, there are 80 TFLite models and 9 TensorFlow models, and most are fine-tuned based on MobileNet V1 or V2. Although DeepPayload claims that it can attack TensorFlow and TFLite models, it only describes how to attack TensorFlow models, and we do not find the latter's implementation. In our experimental process, we first need to convert the TFLite model to a TensorFlow model, perform the attack, and then convert it back. Specifically, to convert a ``.tflite'' model to a ``.pb'' model, we need to use the FlatBuffers compiler (flatc) and a schema file (schema.fbs), which is a FlatBuffers schema that defines the structure of the data in the ``.tflite'' file. By utilizing flatc, the ``.tflite'' model is first parsed according to the schema, enabling the extraction and translation of its structure and data into a format compatible with ``.pb''. Through observations during the experiment, we find that model conversion does not affect the normal performance of the model and the ASR.

\subsubsection{Attack Effectiveness Evaluation on Real-world models}
Finally, we identify 11 real-world models with explicit model information and output category labels, as shown in Table \ref{RQ3_model}. We then collect sufficient image data based on these category labels. As introduced in Section \ref{RQ1}, we implement BARWM and the three baseline backdoor attacks on these models and compare the BA and ASR after the attacks. The evaluation results are listed in Table \ref{RQ3_results}, from which we can draw the following observations and conclusions:
\begin{enumerate}[label=(\roman*)]
	\item For the normal performance of models, the BA values of the backdoor models after BARWM, BadNets, and Invisible Attack are noticeably closer to those of the normal models. The average BA of BARWM is 3.25\% higher than that of DeepPayload and also marginally higher than those of BadNets and Invisible Attack. On more than half of the real-world models, the BA of the backdoor models exceeds that of the normal models (i.e., those underlined in Table \ref{RQ3_results}). The possible reason is that when we implement the backdoor attacks, the victim models also capture the features of many benign samples from the images we collected. 
 
    \item For the backdoor performance, BARWM, BadNets, and Invisible Attack significantly outperform DeepPayload. Specifically, BARWM achieves an average of 12.50\% higher ASR compared to DeepPayload on these real-world models. Furthermore, the ASR of BARWM $>$ 86.39\% across all models, indicating its strong attack stability. Although DeepPayload also achieves relatively effective attacks with a maximum ASR of 91.88\%, it never surpasses BARWM and struggles to maintain stable attack efficiency. Thus, BARWM is more robust than DeepPayload. Compared to BadNets and Invisible Attack, BARWM achieves a comparable ASR overall, with slight outperformance in average ASR, despite variation in performance across different models and datasets. This is sufficient to fully demonstrate the effectiveness and robustness of BARWM. 
    
	\item Considering both normal and backdoor performance, BARWM is a more effective and robust backdoor attack compared to DeepPayload on the real-world models, and performs on par with or better than two typical backdoor attack methods.
\end{enumerate}

\begin{tcolorbox}
	\textbf{Answer to RQ3:} BARWM achieves significantly higher ASR and superior attack stability than DeepPayload while better maintaining the normal performance of the models. Furthermore, it is on par with or even superior to the typical backdoor attack methods, BadNets and Invisible Attack. This indicates that BARWM is more effective and robust on real-world models.
\end{tcolorbox}
\section{Discussion}
\subsection{Attack Characteristics} From the perspective of attack methodology, BARWM employs sample-specific, imperceptible backdoor triggers, which bear a resemblance to perturbations in adversarial attacks \cite{smartapp}. In BARWM, minor perturbations that are imperceptible to humans are generated as backdoor triggers, and these triggers are tailored to specific samples, similar to adversarial perturbations. On one hand, this approach of embedding a hidden backdoor addresses the limitations of adversarial attacks, such as their typically low ASR, thereby exacerbating the security threats posed to real-world DL models. On the other hand, for backdoor attacks, this method enhances stealthiness and efficacy, making them more insidious and challenging to detect and mitigate.

\subsection{Attack Applicability} Our research primarily focuses on backdoor attacks targeting real-world models used in image classification tasks. However, the underlying principles of these attacks are equally applicable to object detection models. In our analysis, we observed that a significant proportion, 34.83\% (31/89), of the models are used for object detection, and the backdoor attack techniques can be seamlessly extended to these models. 

The attack results of backdoor attacks on object detection models include: 

\begin{itemize}
    \item Incorrect detection of target objects. When the trigger is present, the model may detect non-existent objects, thereby generating false positives.
    \item Missed detection of target objects. The presence of the trigger can cause the model to fail to detect actual objects, leading to false negatives.
    \item Missed detection of target objects. The presence of the trigger can cause the model to fail to detect actual objects, leading to false negatives.
    \item Incorrect localization of target objects. The trigger can result in erroneous placement of bounding boxes, affecting the accuracy of object localization.
\end{itemize}

These scenarios demonstrate the broad applicability and potential severity of backdoor attacks across different DL model tasks, highlighting the necessity for robust defensive mechanisms.

\subsection{Model Understanding} Besides extracting DL models, it is crucial to figure out the specific tasks and output information (e.g., category labels) of these models, whether for backdoor attacks or adversarial attacks. Otherwise, the attacks would have no practical significance. In this work, we do not delve into inferring the category labels of the models. Future research can address this gap by employing dynamic analysis techniques. For instance, researchers could use large-scale input testing, where the model is run with images of known categories, and the outputs are observed to infer the category labels. Additionally, debugging the application and scrutinizing inference logs can provide valuable insights into the category labels. These approaches can significantly enhance the understanding of model behaviors and facilitate more effective backdoor and adversarial attacks.

\subsection{Defence of Attack} Our method indeed exacerbates the security threat to real-world models by enhancing the stealthiness and effectiveness of backdoor attacks. This should also serve as a critical call to action for security researchers and developers to enhance their defensive measures. However, for mobile apps, which are complex real-world software systems, the threat might be mitigated through several strategies aimed at preventing model theft. These strategies include: 
\begin{itemize}
    \item Model encryption. On-device models can be encrypted, with the ciphertext stored locally. Mobile apps need to decrypt the ciphertext and load the model into memory for use.
    \item Identity authentication. Models are only granted to specific users for use. Some apps authenticate whether the current user owns a valid authorization token, which is generally distributed during app execution by a remote server.
    \item Model packing. Models can be converted to native C++ code that is much harder to parse. This means one model is stored as native code rather than a plain file.
    \item Weights protection. Model weights can be masked during deployment and unmasked during runtime.
    \item Label encryption. The label file can be encrypted to prohibit attackers from understanding model information and functionality.
\end{itemize}
\section{Conclusion}
In this work, we propose a novel backdoor attack approach on real-world DL models, BARWM, which generates sample-specific triggers and perceptually invariant backdoor samples through DNN-based steganography. We confirm that DNN-based steganography is effective in backdoor attacks. Furthermore, we evaluate the attack stealthiness, and the experimental results demonstrate that BARWM is more stealthy as it does not require changing the model structure and backdoor triggers are imperceptible. Ultimately, on well-understood real-world models, extensive experimental results demonstrate that BARWM is more effective and robust than the baseline methods, which indicates that BARWM will pose a greater security threat.

\bibliographystyle{IEEEtran}
\bibliography{IEEEabrv}

\end{document}